\documentclass[intlimits,twoside,a4paper]{article}

\usepackage[cp1251]{inputenc}

\usepackage[eqsecnum]{cmpj3}
\usepackage{cleveref}
\usepackage{etoolbox}
\usepackage{amsmath}
\usepackage{eqnarray}
\usepackage{soul}
\usepackage{xfrac}

\issue{2022}{25}{3}{33501}
\doinumber{10.5488/CMP.25.33501}

\title[Chaos sync. in a BEC system using FLC]%
{Chaos synchronization in a BEC system using  fuzzy logic controller%
}
\author[E. Tosyali, Y. Oniz, F. Aydogmus]{E. Tosyali\orcid{0000-0001-9118-851X}\refaddr{label1}\thanks{Corresponding author: \email{eren.tosyali@bilgi.edu.tr}.}, 
        Y. Oniz\orcid{0000-0002-8337-7852}\refaddr{label2},
        F. Aydogmus\orcid{0000-0003-1434-2143}\refaddr{label3}}
\addresses{
\addr{label1} Opticianry, Vocational School of Health Services, Istanbul Bilgi University, Kustepe, Sisli, Istanbul, 34387, Turkey
\addr{label2} Faculty of Engineering and Natural Sciences, Department of Mechatronics Engineering, Istanbul Bilgi University, Eyup, Istanbul, 34060, Turkey
\addr{label3} Faculty of Science, Physics, Istanbul University,  Vezneciler, Istanbul, 34134,  Turkey
}


\date{Received April 13, 2022, in final form August 31, 2022}

\begin{document}

\maketitle

\begin{abstract}
Since the presence of chaos in Bose-Einstein condensate (BEC) systems plays a destructive role that can undermine the stability of the condensates, controlling the chaos is of great importance for the creation of the BEC. In this paper, a fuzzy logic controller (FLC) to synchronize the chaotic dynamics of two identical master-slave BEC systems has been proposed. Unlike the conventional approaches, where expert knowledge is directly used to construct the fuzzy rules and membership functions, the fuzzy rules have been constructed using Lyapunov stability {theorem} ensuring the synchronization process. The effectiveness of {the} proposed controller {has been} demonstrated numerically.

\printkeywords 
%
\end{abstract}

\section{Introduction}
\label{sec1}

 {Bose-Einstein condensation (BEC) is a process, in which the system forms a single coherent matter wave after the temperature of boson gases is reduced below a critical level. 
The theoretical background of BEC was set by Einstein in}~\cite{Work2,Work3},  {with the idea that the boson gases will experience a phase transition at their critical temperature,}  {whereas the idea was experimentally verified in 1995 using the dilute atomic vapor of rubidium and sodium~\cite{Anderson1,Davis1}}.

{Despite the fact that} the temperatures obtained with lasers  are quite  {low}, to be able to form BEC, an additional cooling method  {is required to enable the} atoms with relatively higher energy to escape from the trap~\cite{KETTERLE1996181}. In this  {cooling method, which} reduces the kinetic energy of the entire system,  {the}  {m}agneto-{o}ptical  {t}rap and lasers are turned off  {while} another magnetic field is activated at the same time. The energy of the atoms at the center of the trap is  {considerably} smaller than the energy of the atoms at the corners of the trap. Trapped  {dilute} boson gases interact  {with each other} due to their physical properties or due to collisions. In the interacting gases, only weakly interacting states caused by binary collisions ($s$-wave scattering) are considered,  {as} it is not possible to express the system macroscopically with a single wave function in non-weak interactions.

Radiofrequency is used to enable atoms with higher energies to escape from the trap, which provides a change  {in} the  {spinning} direction of the atoms.  {This process generates } a repulsive force for atoms, where the magnetic field and  {the} magnetic moment are parallel.  {An attractive force occurs among the atoms due to opposite magnetic moments}. The repulsive force separates the atomic cloud as trapped and untrapped, and allows the atoms with more energy standing  {at} the corners to be thrown out of the trap. The atoms in the trap collide and transfer their momentum to each other; and they come into equilibrium at a new low thermal energy called back thermalization. This process is repeated until the critical temperature  {is reached}~\cite{WorkC}.

The condensation of weakly interacting boson gases, for which {the temperature is close to zero}, is {well} expressed by the Gross-Pitaevskii equation (GPE). This equation was derived in 1961 by Gross and Pitaevskii  independently and with different techniques to describe weakly interacting  {dilute} boson gases~\cite{Work9,Work10}. Basically, the GPE is the nonlinear Schr\"{o}dinger equation derived from mean-field theory. The GPE gives favorable results  {in} the experiments  {with} weakly interacting BEC. 

It is of great importance to study  {how to control} the chaos of a BEC system in an optical lattice which exhibits many rich and complex phenomena typical of nonlinear systems~\cite{eren1,eren2,eren3,eren4,eren5}. An important approach to consider is the synchronization problem from a control theory perspective. 

Different control schemes including feedback control~\cite{idowu2013synchronization,guo2017projective,mobayen2018synchronization,ZhangZhiYing110503}, sliding mode control,~\cite{chen2012chaotic,vaidyanathan2015hybrid} and fuzzy logic control~\cite{yau2008chaos,vaidyanathan2016takagi} have been proposed over the last decade for the synchronization problem of chaotic systems. The main drawback of the feedback control schemes is that the control signals are generated relying on the mathematical model of the chaotic system. However, in many applications the dynamics of the system will be perturbed because of the uncertainties in the system parameters and external disturbances. Hence, these controllers may fail to provide reliable results.
The sliding mode  {control} approach can be pointed out among the most effective robust controllers to handle high-order nonlinear systems. However,  {this approach} inherently suffers from the chattering problem. On the other hand, fuzzy logic controllers (FLC) provide an easy but effective way to cope with uncertain and nonlinear system dynamics, and they were successfully applied in many areas such as control~\cite{topalov2011neuro,khanesar2015direct}, decision making~\cite{tirkolaee2020novel,blanco2017fuzzy}, prediction~\cite{kalaiarassan2018one,saez2014fuzzy}, forecasting~\cite{cheng2016fuzzy,atsalakis2019bitcoin}, and modelling~\cite{kostikova2016expert,aghbashlo2017fuzzy}. Recently, fuzzy logic control of chaotic systems has become an active research area. In the fuzzy logic control, the output of the controller is determined using the fuzzy inference. The rules typically  {rely} on  expert knowledge. Although promising results have been reported in the literature for the use of this conventional scheme in chaos synchronization~\cite{bulut2019fuzzy,chou2013fuzzy}, the performance of these controllers might significantly degrade if expert knowledge is incomplete and/or uncertain. To alleviate this issue, adaptive approaches are commonly preferred in the design of FLCs, in which the controller parameters are updated to lead the synchronization error to zero ~\cite{wang2020fuzzy,kuo2007design}. Despite the fact that the adaptive schemes can provide fairly good results, their time requirements for the adaptation process might pose a problem in real-time applications. In the proposed work,  {the} Lyapunov stability theorem was directly employed to construct the consequent part of the fuzzy rules such that two identical master-slave BEC systems can be synchronized. One of the most prominent advantages of this control scheme is that the  stability in the Lyapunov sense of error dynamics of two identical chaotic BEC motions was ensured. The feasibility and effectiveness of the proposed controller were demonstrated by numerical simulation results.

\section{Description of system }
\label{sec2}

GPE including   {macroscopic} wave function can well describe the evolution of the BEC  {simultaneously with regard to } time and space~\cite{Work9,Work10}.  {One-dimensional (1D)} GPE  {can be described} as below:
\begin{equation}
\ri \hbar \frac{\partial }{\partial t}\Psi \left( x,t\right) =-\frac{\hbar ^{2}%
}{2m}\frac{\partial ^{2}}{\partial x^{2}}\Psi \left( x,t\right) +\left[
V_{\rm{ext}}\left( x\right) +g_{1D}\left\vert \Psi \left( x,t\right) \right\vert
^{2}\right] \Psi \left( x,t\right) ,  \label{eq1}
\end{equation}
{where} $m$  {stands} for the mass of the atoms which  {constitute} the BEC, $V_{\rm{ext}}$ is  {the} external potential with  {tilted} term trapping  {from} the BEC, and $g_{1D}$ is  {the} one dimensional interaction term between  {the} atoms  { defined as:}
 \begin{equation} 
 g_{1D}=\frac{g_{3D}}{2\piup a_{r}^{2}}=2a_{s} \hbar \omega_{r}, \nonumber  
 \end{equation}
{with} $a_{s}$  {being} the $s$-wave scattering length between atoms. $s$-wave scattering length could be positive or negative depending on the interactions whether  it is repulsive or attractive, respectively. In our case, its value is negative due to attractive interactions. $\omega_{r}$ is  {the} ground state of a harmonic frequency of the oscillator.

\begin{figure}
\centerline{\includegraphics[width=0.45\textwidth]{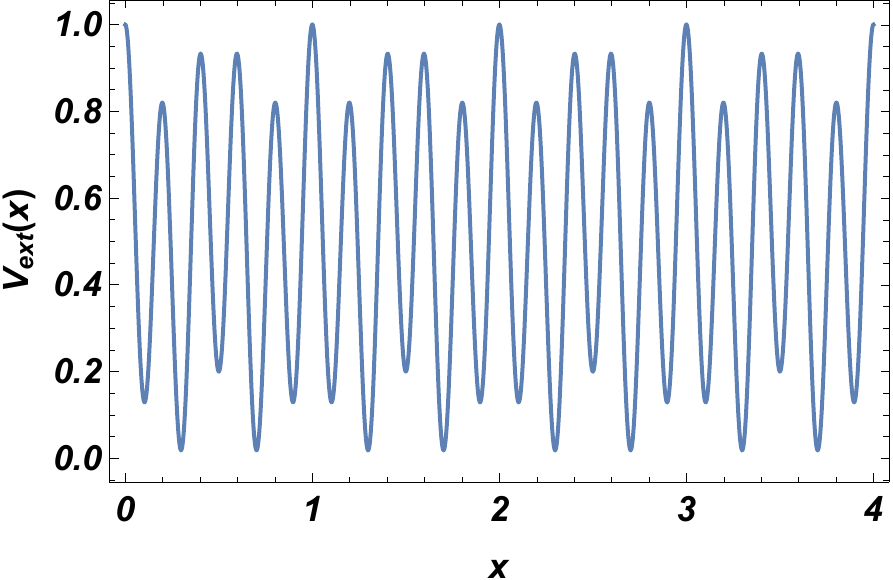}   \hspace{0.2cm}\includegraphics[width=0.45\textwidth]{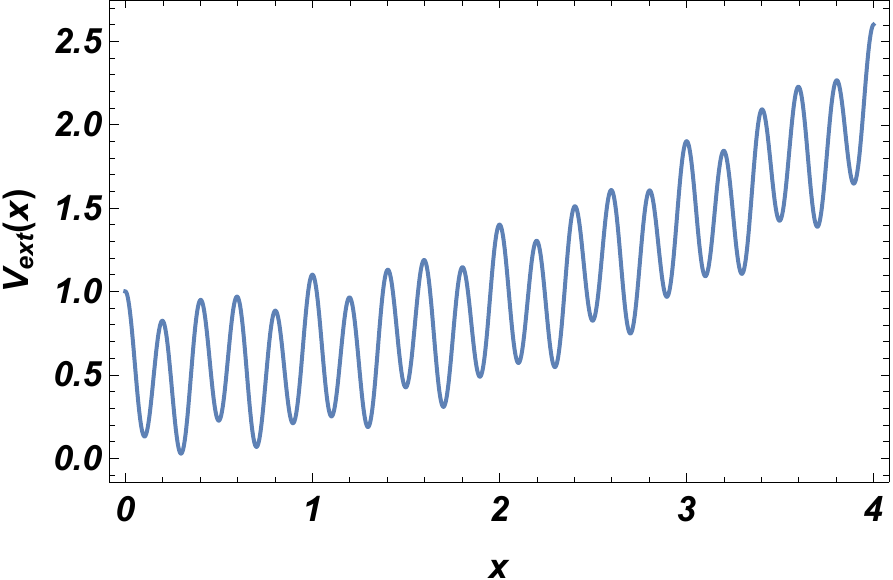}  }
\begin{center} (a) \hspace{6cm} (b) \end{center}
\caption{(Colour online) Plot of the bichromatic optical lattice potential with the parameters $\nu_{1}=1$, $\nu_{2}=0.8$, $\omega_{1}=2\piup$, $\omega_{2}=5\piup$, (a) $ {F}=0$, (b) $ {F}=0.1$. }
\label{fig_1}
\end{figure}

The external trap potential $V_{\rm{ext}} \left(x \right )$ is given as:
\begin{equation}
V_{\rm{ext}} \left(x \right )=V_{1}\cos^{2}(\omega_{1} x)+V_{2}\cos^{2}(\omega_{2} x)+Fx. \label{eq2}
\end{equation}
$V_{\rm{ext}}$  {comprises} two parts: while the first part of double well potential with two  {frequencies} is related to  {the} optical lattice potential, the second part is related to  {the} tilted potential. Here, $V_{1}$ and $V_{2}$ are the amplitudes,  $F$ is  {the} internal  {f}orce and  $Fx$  {corresponds to} the tilted potential, which makes  {the} atoms tunnelling out from the potential and accelerates  {them} in the $x$ direction.  {1D Hamiltonian $\left(H_{F}=H_{0}+Fx \right)$ tends to infinity when   $\vert x \rvert \rightarrow \infty$~\cite{neciu1991}. However, the  {H}amiltonian is always bounded if the lattice size is finite $-L\leqslant x \leqslant L$~\cite{bond_2003}. The BEC system of this study is bounded with 100 lattice sites. The number of lattice sites was determined empirically as $100$, which implies that $L\sim 100\piup k^{-1}$ ($k={2\piup}/{850}$~nm$^{-1}$)~\cite{fallani2004,Denschlag_2002,FANG200561}. To meet the requirements on the numbers of the lattice sites and thus on the boundary conditions, the simulated studies were carried out for 1000 steps with a step size of 0.1, and for 10000 steps with a step size of 0.01.}  In figure~\ref{fig_1}, the evolution of the external potential for parameters set $\nu_{1}=1$, $\nu_{2}=0.8$, $\omega_{1}=2\piup$, $\omega_{2}=5\piup$, (a) $ {F}=0$, (b) $ {F} =0.1$  {is illustrated.}

\begin{figure}
\centerline{\includegraphics[width=0.5\textwidth]{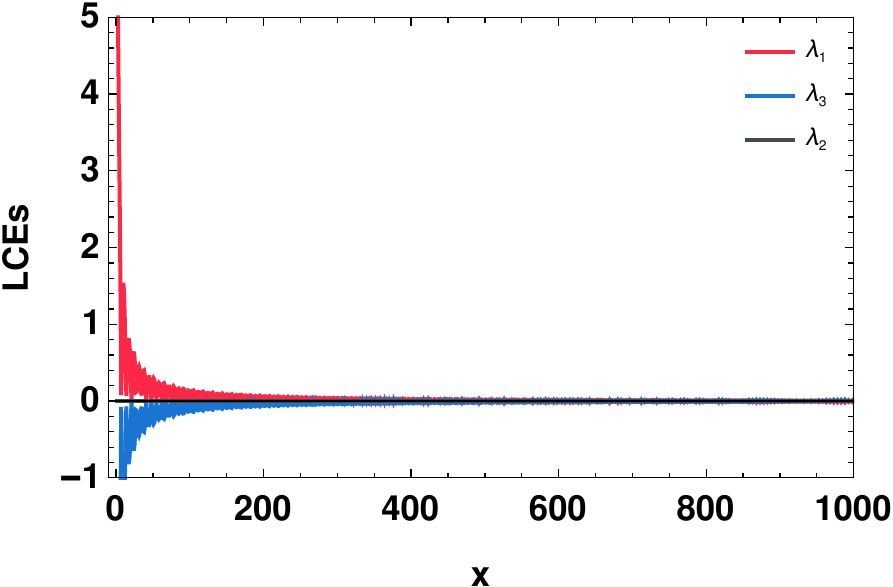} }
\vspace*{0pt}
\caption{(Colour online) LCEs for paramter sets: $J=0.4$, $\nu_{1}=1$, $\nu_{2}=0.8$, $\omega_{1}=2\piup$, $\omega_{2}=5\piup$, $\Gamma=0.1$, $\eta=-0.015$, $\gamma=0.5$ and initial condition: $(x_{1}=1$, $y_{1}=-1)$. }
\label{fig1aa}
\end{figure}

 {There are different time-dependent ansatzs to solve the GPE~\cite{chua2006,hai2009}. In this study, the following widely-used form of the time-dependent wave function is preferred:}
\begin{equation}
\Psi \left( x,t\right) =\Phi \left( x\right) \re^{{-\ri\mu t}/{\hbar }},
\label{eq3}
\end{equation}
here, $\mu$ is the chemical potential of the condensate and $\Phi(x)$  is a real function independent of time. Normalized $\Phi(x)$ gives the total number of particles in the system, i.e.,
\begin{equation}
\int \left\vert \Phi \left( x\right) \right\vert ^{2}\rd x=N, \label{eq4}
\end{equation}
where $N$ is the particle number. Substitution of equations~\eqref{eq2} and~\eqref{eq3}  into equation~\eqref{eq1}   {yields}:
\begin{equation}
\mu \Phi \left(x \right)=-\frac{\hbar ^{2}}{2m}\frac{\rd ^{2}}{\rd  x^{2}}\Phi \left( x\right) +\left[
V _{1}\cos^{2}\left( \omega_{1} x\right)+V_{2}\cos^{2}(\omega_{2} x)+Fx+g_{1D}\left\vert \Phi \left( x\right) \right\vert
^{2}\right] \Phi \left( x\right).  \label{eq5}
\end{equation}
 {Using the dimensionless parameters} $\upsilon_{1}={2mV_{1}}/{\hbar ^{2}}$, $\upsilon_{2}={2mV_{2}}/{\hbar ^{2}}$, $\gamma ={2m\mu }/{\hbar ^{2}}$, $\eta ={2mg_{0}}/{\hbar ^{2}}$, $\Gamma={2mF}/{\hbar^2}$,  {the equation}~\eqref{eq5}  {can be written as:}
\begin{equation}
\frac{\rd^{2}\Phi }{\rd x^{2}}=\left[ \upsilon _{1}\cos^{2}\left( \omega_{1} x\right)  +\upsilon _{2}\cos^{2}\left( \omega_{2} x\right)+\Gamma x -\gamma +\eta \left\vert \Phi
\right\vert ^{2}\right] \Phi.  \label{eq6}
\end{equation}
 {The solution of equation}~\eqref{eq6}  {has the following form:} 
\begin{equation}
\Phi(x)=\phi(x) \re^{\ri\theta(x)}, \label{eq7}
\end{equation}
 where $\phi$ and $\theta$ are real functions of $x$, expressing  {the} amplitude and  {the} phase, respectively. The first derivative of the phase is proportional to the velocity field, and the squared amplitude corresponds to the density of the atoms in the condensate. {Substituting equation}~\eqref{eq7}  {into equation}~\eqref{eq6} results in two coupled equations, which correspond to the real and imaginary parts.
\begin{equation}
\frac{\rd^{2}\phi }{\rd x^{2}}=\phi \left( \frac{\rd \theta }{\rd x}\right) ^{2} +\left[
\upsilon _{1}\cos^{2}\left( \omega_{1} x\right)+ \upsilon _{2}\cos^{2}\left( \omega_{2} x\right)+\Gamma x-\gamma +\eta \left\vert \phi \right\vert ^{2}\right]
\phi ,  \label{eq8a}
\end{equation}
\begin{equation}
\frac{\rd^{2} \theta}{\rd x^{2}}+2\frac{1}{\phi} \frac{\rd \theta}{\rd x} \frac{\rd \phi}{\rd x} =0. \label{eq8b}
\end{equation}
 {Integration of equation~\eqref{eq8b} gives the relation between the velocity field and the particle number density,}
\begin{equation}
J=2\phi^{2}\left(\frac{\rd \theta}{\rd x}\right), \label{eq9}
\end{equation}
{where $J$ represents the steady super-fluidity phase in fluid dynamics.}  {Using} $J$  {in equation}~\eqref{eq8a},  {the following  nonlinear equation can be obtained:}
\begin{equation}
\frac{\rd^{2}\phi }{\rd x^{2}}=\frac{J^{2}} {4\phi ^{3}}+\left[
\upsilon _{1}\cos^{2}\left( \omega_{1} x\right)  +\upsilon _{2}\cos^{2}\left( \omega_{2} x\right)+\Gamma x-\gamma +\eta \left\vert \phi \right\vert ^{2}\right]
\phi .\label{eq10}
\end{equation}
The equation~\eqref{eq10}  {was} solved numerically for $1000$ steps with $100$ eliminated steps,  {with a step size of} $0.1$  {to satisfy the 100 lattice site boundary condition. In the simulation, the following values for the parameters were assumed}: $J=0.4$, $\nu_{1}=1$, $\nu_{2}=0.8$, $\omega_{1}=2\piup$, $\omega_{2}=5\piup$, $\Gamma=0.1$, $\eta=-0.015$, $\gamma=0.5$ and possible initial conditions $(x_{1}=1$, $y_{1}=-1)$ by the Runge-Kutta method.  {The} Lyapunov characteristic exponents (LCEs) of  {the} BEC system given in figure~\ref{fig1aa} are $\lambda_{1}=0.0046543$, $\lambda_{2}=0$ and $\lambda_{3}=-0.00465435$.  {For the  given system parameters and initial conditions, the system exhibits chaotic behaviour due to the presence of positive LCE}.

{Using} the transformations ${\rd \phi}/{\rd x}={\rd x_{1}}/{\rd x}$ and ${\rd ^{2}\phi}/{\rd x^{2}}={\rd y_{1}}/{\rd x}$, {the} equation~\eqref{eq10}  {can be re-written with the following}  first-order coupled equations, which  defines the master system as:
\begin{equation}
\frac{\rd x_{1}}{\rd x}=y_{1}, \label{eq11a}
\end{equation}
\begin{equation}
\frac{\rd y_{1}}{\rd x}=\frac{J^{2}} {4x_{1} ^{3}}+\left[
\upsilon _{1}\cos^{2}\left( \omega_{1} x\right)  +\upsilon _{2}\cos^{2}\left( \omega_{2} x\right)+\Gamma x-\gamma +\eta \left\vert  x_{1} \right\vert ^{2}\right] x_{1}, \label{eq11b}
\end{equation}
{and the dynamics of the} slave system  {are given below}:
\begin{equation}
\frac{\rd x_{2}}{\rd x}=y_{2}. \label{eq3_2a}
\end{equation}
\begin{eqnarray}
\frac{\rd y_{2}}{\rd x}=\frac{J^{2}} {4x_{2} ^{3}}+\left[
\upsilon _{1}\cos^{2}\left( \omega_{1} x\right)  +\upsilon _{2}\cos^{2}\left( \omega_{2} x\right)+\Gamma x-\gamma +\eta \left\vert  x_{2} \right\vert ^{2}\right] x_{2}+ \Delta f(x_{2},y_{2})+u(x),  \label{eq3_2b}
\end{eqnarray}
where $u$ is  {the} control input, $\Delta f(x_{2},y_{2})$ is the uncertain term which is assumed bounded, i.e., 
$\Delta f(x_{2},y_{2})\leqslant \alpha$,
where $\alpha$ is a positive constant. It is also assumed that $\Delta f(x_{2}, y_{2})$ satisfy all the necessary conditions, such as system~\eqref{eq3_2a},~\eqref{eq3_2b} having  {a} unique solution in the spatial evolution $x_{0}+\infty$, for any given initial condition.
In this paper, the control input $u(x)$  {is derived} to synchronize  {the master and slave} system.

\section{Fuzzy design for chaotic synchronization }
\label{sec3}
\begin{figure}[h]
	\centerline{\includegraphics[width=0.5\textwidth]{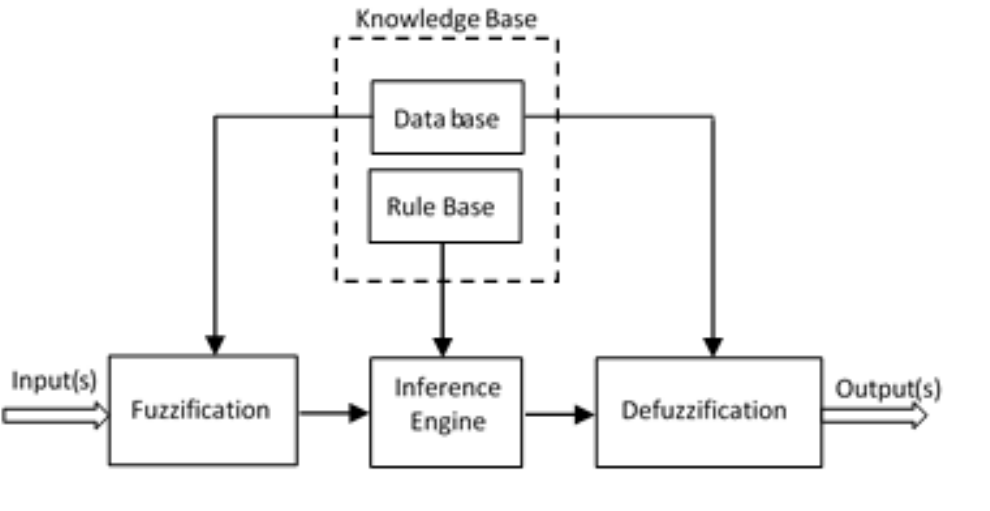}     }
	\vspace*{0pt}
	\caption{Structure of a fuzzy inference system.}
	\label{fig1a}
\end{figure}

Fuzzy inference can be defined as a process, in which the given input(s)  are mapped to the output(s) using the fuzzy set theory. Figure~\ref{fig1a}   {depicts the general structure of a fuzzy inference system, which consists of four main function blocks}: 
\begin{enumerate}
\item  The knowledge base comprises expert knowledge and it can be separated into two parts: database and rule base. In the database, the type and parameters of the membership functions are stored, whereas the rule base includes the fuzzy ``if-then'' rules.
\item In the fuzzification stage, the crisp input variables are transformed into fuzzy sets using the expert knowledge stored in the database. Hence, the outputs of this block are fuzzy sets.
\item In the inference engine, using the fuzzy ``if-then'' rules stored in the rule base, rule consequents are computed for each rule and aggregated into a single fuzzy output set.
\item Output fuzzy sets are transformed into crisp output values in the defuzzification step.
\end{enumerate}

\begin{figure}[ht]
\centerline{\includegraphics[width=0.55\textwidth]{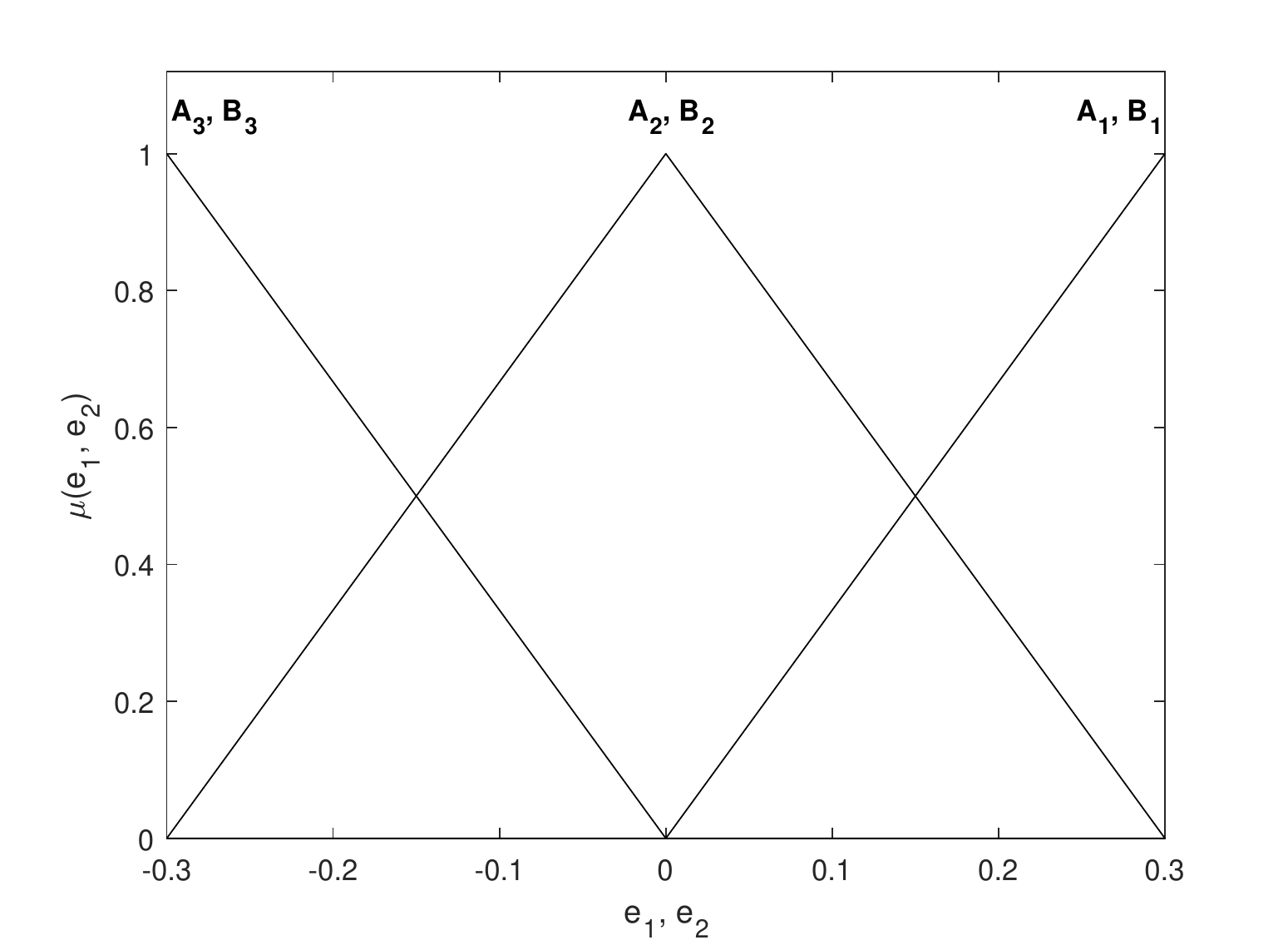} }    
\vspace*{0pt}
\caption{Membership functions. }
\label{fig2a}
\end{figure}

In order to synchronize  {the} master and slave system with the dynamics given in  equations~\eqref{eq11a}--\eqref{eq3_2b},  {the} error states are defined as $e_{1}=x_{2}-x_{1}$ and $e_{2}=y_{2}-y_{1}$  {by s}ubtracting  {the} master system from the slave system.  {The} error dynamics  {can be specified as}:
\begin{equation}
\frac{\rd e_{1}}{\rd x}=e_{2},
\label{eq4a}
\end{equation}
\begin{eqnarray}
\frac{\rd e_{2}}{\rd x}=\left[
\upsilon _{1}\cos^{2}\left( \omega_{1} x\right)  +\upsilon _{2}\cos^{2} \left( \omega_{2} x\right)+\Gamma x-\gamma \right]e_{1} +\Delta f(x_{2},y_{2})+u_{L}.
\label{eq4b}
\end{eqnarray}
 {T}he control input $u(x)$   {in equation}~\eqref{eq3_2b}  {can be written as} $u(x)= {u}_{eq}+u_{L}$,  {where} $u_{eq}=-\eta \left\vert  x_{2} \right\vert ^{2} x_{2}+\eta \left\vert  x_{1} \right\vert ^{2} x_{1}$.  A fuzzy logic controller is proposed to construct the control input  $u_{L}$, where the error dynamics in equation~\eqref{eq4a} and~\eqref{eq4b} are utilized as the input signals of the FLC. These incoming signals, $(e_{1},e_{2})$, are fuzzified using triangular membership functions shown in figure~\ref{fig2a}, and associated with $A_i$ and $B_j$ fuzzy subsets, respectively, which are defined by their corresponding membership functions $\mu_{A_i}(e_1)$ and $\mu_{B_j}(e_2)$ for $i=1,\dots,I$ and $j=1,\dots,J$. 

The fuzzy ``if-then'' rule $R_{ij}$ with the inputs $(e_{1},e_{2})$ and the output $u_L$ can be stated as:
\begin{center}
   $R_{ij}$: If $e_1$ is  $A_i$ and $e_2$ is $B_j$, then $u_L=u_{L,ij}(e_{1},e_{2})$, 
\end{center}
where $u_{L,ij}$  {is an} analytical function of $e_{1},e_{2}$  {that} stabilizes the error dynamics in equation~\eqref{eq4a} and~\eqref{eq4b}. Minimum t-norm is employed to calculate the firing strengths of  {each} rule, i.e., $ W_{ij}=\min [\mu_{A_i}(e_1), \mu_{B_j}(e_2)]$, whereas   {t}he centroid defuzzification method is used to compute the output control signal $u_{L}$:
\begin{equation}
u_{L}=\frac{\sum_{i=1}^{I}\sum_{j=1}^{J}{W}_{ij}u_{L,ij}}{\sum_{i=1}^{I}\sum_{j=1}^{J}{W}_{ij}}.
\label{eq4d}
\end{equation}
The fuzzy rules used in this study are presented in table~\ref{tab1}, in which $e_{1}$ and $e_{2}$  correspond to the input variables used in the antecedent part of the rules, and $u_{L_{ij}}$  \  {denotes} the output variable of the consequent. As illustrated in figure~\ref{fig2a}, three membership functions are used for each input: $A_1$, $A_2$ and $A_3$ for the input signal $e_1$, and $B_1$, $B_2$ and $B_3$ for the input signal $e_2$. The membership functions $\{A_1, B_1\}$, $\{A_2, B_2\}$ and $\{A_3, B_3\}$  {correspond to} positive, zero and negative error dynamics,  respectively.

\begin{table}
\caption{Rule Table of FLC.}
\label{tab1}
\begin{tabular*}{\textwidth}{@{}l*{15}{@{\extracolsep{\fill}}l}}
\hline
Rule&\multicolumn{2}{l}{Antecedent } &Consequent\\
\hline
&$e_{1}$ & $e_{2}$ & $u_{L,ij}$ \\
\hline
 1 & $A_1$ & $B_1$ & $u_{L,11}$ \\

 2 & $A_1$ & $B_2$ & $u_{L,12}$ \\

 3 & $A_1$ & $B_3$ & $u_{L,13}$ \\

 4 & $A_2$ & $B_1$ & $u_{L,21}$ \\

 5 & $A_2$ & $B_2$ & $u_{L,22}$ \\

 6 & $A_2$ & $B_3$ & $u_{L,23}$ \\
 
 7 & $A_3$ & $B_1$ & $u_{L,31}$ \\
  
 8 & $A_3$ & $B_2$ & $u_{L,32}$ \\
    
 9 & $A_3$ & $B_3$ & $u_{L,33}$ \\
 \hline
\end{tabular*}
\end{table}

 {The following function, which is positive and continuously differentiable, is selected as the Lyapunov function candidate}:
\begin{equation}
V=\frac{1}{2}\left(e_{1}^{2}+e_{2}^{2}\right).
\label{eq4e}
\end{equation}
To ensure the Lyapunov stability, the following condition should be met~\cite{Work31}:
\begin{equation}
\dot{V}=e_{1}\dot{e}_{1}+e_{2}\dot{e}_{2}<0,
\label{eq4f}
\end{equation}
 {which requires} $\dot{e}_{2}<-{e_{1}\dot{e}_{1}}/{e_{2}}$.  {For t}he following cases,  {the consequents of the FLC are proposed such that the} stability condition  {is ensured}.

\textbf{Case 1:}  $e_{2}<0$

For $e_{2}<0$,  {the stability condition requires that:}
\begin{equation}
\dot{e}_{2}>-e_{1}.
\label{eq4g}
\end{equation}
Substituting equation~\eqref{eq4g} into equation~\eqref{eq4a} and~\eqref{eq4b} produces
\begin{equation}
\left[\upsilon _{1}\cos^{2}\left( \omega_{1} x\right)  +\upsilon _{2}\cos^{2}\left( \omega_{2} x\right)+\Gamma x-\gamma\right]
e_{1}+\Delta f +u_{L}>-e_{1},
\label{eq4h}
\end{equation}
then $-\left[\upsilon _{1}\cos^{2}\left( \omega_{1} x\right)  +\upsilon _{2}\cos^{2}\left( \omega_{2} x\right)+\Gamma x-\gamma-1\right]
e_{1}-\Delta f <u_{L}$  {should be provided. Assuming} that the uncertainty term $\Delta f$ is bounded, that is $|\Delta f|\leqslant \alpha$,  { with $\alpha$ being a positive constant, the following equality can be derived:}
\begin{equation}
-\left[\upsilon _{1}\cos^{2}\left( \omega_{1} x\right)  +\upsilon _{2}\cos^{2}\left( \omega_{2} x\right)+\Gamma x-\gamma+1\right]
e_{1} +\alpha =u_{1}^{*}.
\label{eq4j}
\end{equation}

\textbf{Case 2:} $e_{2}>0$

 {If} $e_{2}>0$,  {then}:
\begin{equation}
\dot{e}_{2}<-e_{1},
\label{eq4k}
\end{equation}
 {should be provided to meet the stability condition.} Substituting equation~\eqref{eq4k} into equation~\eqref{eq4a} and~\eqref{eq4b}  {results in}:
\begin{equation}
\left[\upsilon _{1}\cos^{2}\left( \omega_{1} x\right)  +\upsilon _{2}\cos^{2}\left( \omega_{2} x\right)+\Gamma x-\gamma\right]
e_{1}+\Delta f +u_{L}<-e_{1}.
\label{eq4l}
\end{equation}
 {Hence, the control input $u_L$ should satisfy the following inequality:}
\begin{equation}
 -\left[\upsilon _{1}\cos^{2}\left( \omega_{1} x\right)  +\upsilon _{2}\cos^{2}\left( \omega_{2} x\right)+\Gamma x-\gamma-1\right]
e_{1}-\Delta f >u_{L}.   
\end{equation}
 {A new term $u_{2}^{*}$ can be derived as:} 
\begin{equation}
-\left[\upsilon _{1}\cos^{2}\left( \omega_{1} x\right)  +\upsilon _{2}\cos^{2}\left( \omega_{2} x\right)+\Gamma x-\gamma+1\right]
e_{1} -\alpha =u_{2}^{*}.
\label{eq4m}
\end{equation}
According to table~\ref{tab1}, $e_{2}$ is negative for Rules 3, 6 and 9,  {which corresponds to the above mentioned \textbf{Case~1}. If the terms in the consequent parts of these rules are selected as $u_{1}^{*}=u_{L,13}=u_{L,23}=u_{L,33}$, then the stability condition will be satisfied, i.e.,  $\dot{V}<0$, and the error states will be asymptotically driven to zero. Similarly, for Rules 1, 4 and 7, $e_{2}$ is positive, which corresponds to \textbf{Case 2}. Hence, if the consequent parameters of these rules are selected as $u_{2}^{*}=u_{L,11}=u_{L,21}=u_{L,31}$, then the stability can be ensured.} 
\\

\textbf{Case 3:} $e_{1}>0$ and $e_2 \in 0$

In table~\ref{tab1} for Rule 2, $e_{1}$ is positive,   {whereas} $e_{2}$ is zero.   {To satisfy the Lyapunov stability condition stated in equation~\eqref{eq4f}, the following equality should be provided}:
\begin{equation}
e_{1}+\dot{e}_{2}=-{\rm{sgn}}(e_{2}),
\label{eq4n}
\end{equation}
where 
$${\rm{sgn}}(e_{2})=\begin{cases}
 1: e_{2}>0, \\
 -1: e_{2}<0.
\end{cases}$$
{As} $e_{1}$ is positive,   {then}
\begin{equation}
\dot{e}_{2}<-{\rm{sgn}}(e_{2}),
\label{eq4o}
\end{equation}
{should be satisfied.}
Substituting equation~\eqref{eq4o} into equation~\eqref{eq4a} and~\eqref{eq4b} yields
\begin{equation}
\left[\upsilon _{1}\cos^{2}\left( \omega_{1} x\right)  +\upsilon _{2}\cos^{2}\left( \omega_{2} x\right)+\Gamma x-\gamma \right]e_{1}+\Delta f +u_{L}<-{\rm{sgn}}(e_{2}).
\label{eq4p}
\end{equation}
The equation can be written as
\begin{equation}
u_{L}<-{\rm{sgn}}(e_{2})-\left[\upsilon _{1}\cos^{2}\left( \omega_{1} x\right)  +\upsilon _{2}\cos^{2}\left( \omega_{2} x\right)+\Gamma x-\gamma \right]e_{1}-\Delta f, 
\label{eq4r}
\end{equation}
 {which implies that if $u_{L,12}$ is selected as:}
\begin{equation}
u_{L,12}=-{\rm{sgn}}(e_{2})-\left[\upsilon _{1}\cos^{2}\left( \omega_{1} x\right)  +\upsilon _{2}\cos^{2}\left( \omega_{2} x\right)+\Gamma x-\gamma \right]e_{1}-\alpha,
\label{eq4s}
\end{equation}
  {then it will satisfy the stability condition.}
\\

\textbf{Case 4:} $e_{1}<0$ and $e_2 \in 0$

The controller $u_{L,32}$ in Rule 8   {can be derived in a way similar to} Rule 2,   {which requires:} 
\begin{equation}
\dot{e}_{2}>-{\rm{sgn}}(e_{2}).
\label{eq4t}
\end{equation}
Substituting equation~\eqref{eq4t} into equation~\eqref{eq4a} and~\eqref{eq4b} yields
\begin{equation}
\left[\upsilon _{1}\cos^{2}\left( \omega_{1} x\right)  +\upsilon _{2}\cos^{2}\left( \omega_{2} x\right)+\Gamma x-\gamma \right]e_{1}+\Delta f +u_{L}>-{\rm{sgn}}(e_{2}).
\label{eq4u}
\end{equation}
The equation can be written as
\begin{equation}
u_{L}>-{\rm{sgn}}(e_{2})-\left[\upsilon _{1}\cos^{2}\left( \omega_{1} x\right)  +\upsilon _{2}\cos^{2}\left( \omega_{2} x\right)+\Gamma x-\gamma \right]e_{1}-\Delta f. 
\label{eq4v}
\end{equation}
  {To ensure the stability, the controller $u_{L,32}$ should be selected as:}
\begin{equation}
u_{L,32}=-{\rm{sgn}}(e_{2})-\left[\upsilon _{1}\cos^{2}\left( \omega_{1} x\right)  +\upsilon _{2}\cos^{2}\left( \omega_{2} x\right)+\Gamma x-\gamma \right]e_{1}+\alpha.
\label{eq4w}
\end{equation}

\textbf{Case 5:} $e_{1} \in 0$ and $e_2 \in 0$

For Rule 5 in Table~\ref{tab1}, the error states $e_{1}$ and $e_{2}$  {are both} zero. Therefore, $u_{L,22}=0$.

The control function of $u_{L_{ij}}$ with $i=1,2,3$ and $j=1,2,3$ depending on $e_{1},e_{2}$ is summarized below.   {The controller parameters are derived such that} all of the rules in the FLC can lead to Lyapunov stable subsystems under the same Lyapunov function~\eqref{eq4e}. 
  {Therefore, it is ensured that the} chaotic master and slave system   {will be} synchronized.
\begin{eqnarray}
u_{L,11}&=&u_{L,21}=u_{L,31}  = -\left[\upsilon _{1}\cos^{2}\left( \omega_{1} x\right)  +\upsilon _{2}\cos^{2}\left( \omega_{2} x\right)+\Gamma x-\gamma+1\right]
e_{1} -\alpha, \nonumber \\
u_{L,13}&=&u_{L,23}=u_{L,33}=-\left[\upsilon _{1}\cos^{2}\left( \omega_{1} x\right)  +\upsilon _{2}\cos^{2}\left( \omega_{2} x\right)+\Gamma x-\gamma+1\right]
e_{1} +\alpha, \nonumber \\
u_{L,12}&=&-{\rm{sgn}}(e_{2})-\left[\upsilon _{1}\cos^{2}\left( \omega_{1} x\right)  +\upsilon _{2}\cos^{2}\left( \omega_{2} x\right)+\Gamma x-\gamma \right]e_{1}-\alpha, \nonumber \\
u_{L,32}&=&-{\rm{sgn}}(e_{2})-\left[\upsilon _{1}\cos^{2}\left( \omega_{1} x\right)  +\upsilon _{2}\cos^{2}\left( \omega_{2} x\right)+\Gamma x-\gamma \right]e_{1}+\alpha, \nonumber \\
u_{L,22}&=&0.
\label{eq3_2bc}
\end{eqnarray}

\section{Simulations and results}
\label{sec5}

\begin{figure}[h!]
 	\centerline{\includegraphics[width=0.45\textwidth]{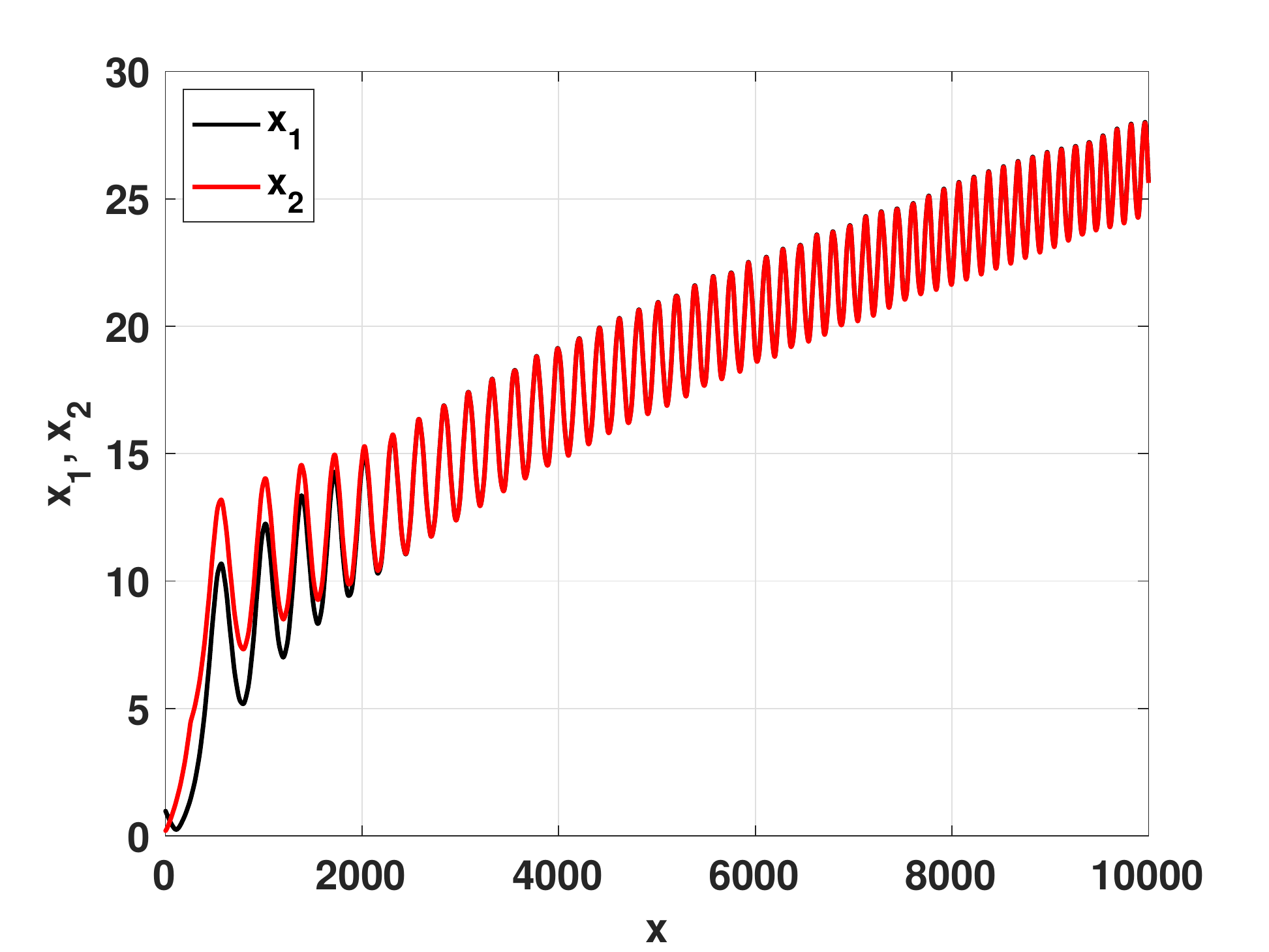}   \hspace{0.2cm}\includegraphics[width=0.45\textwidth]{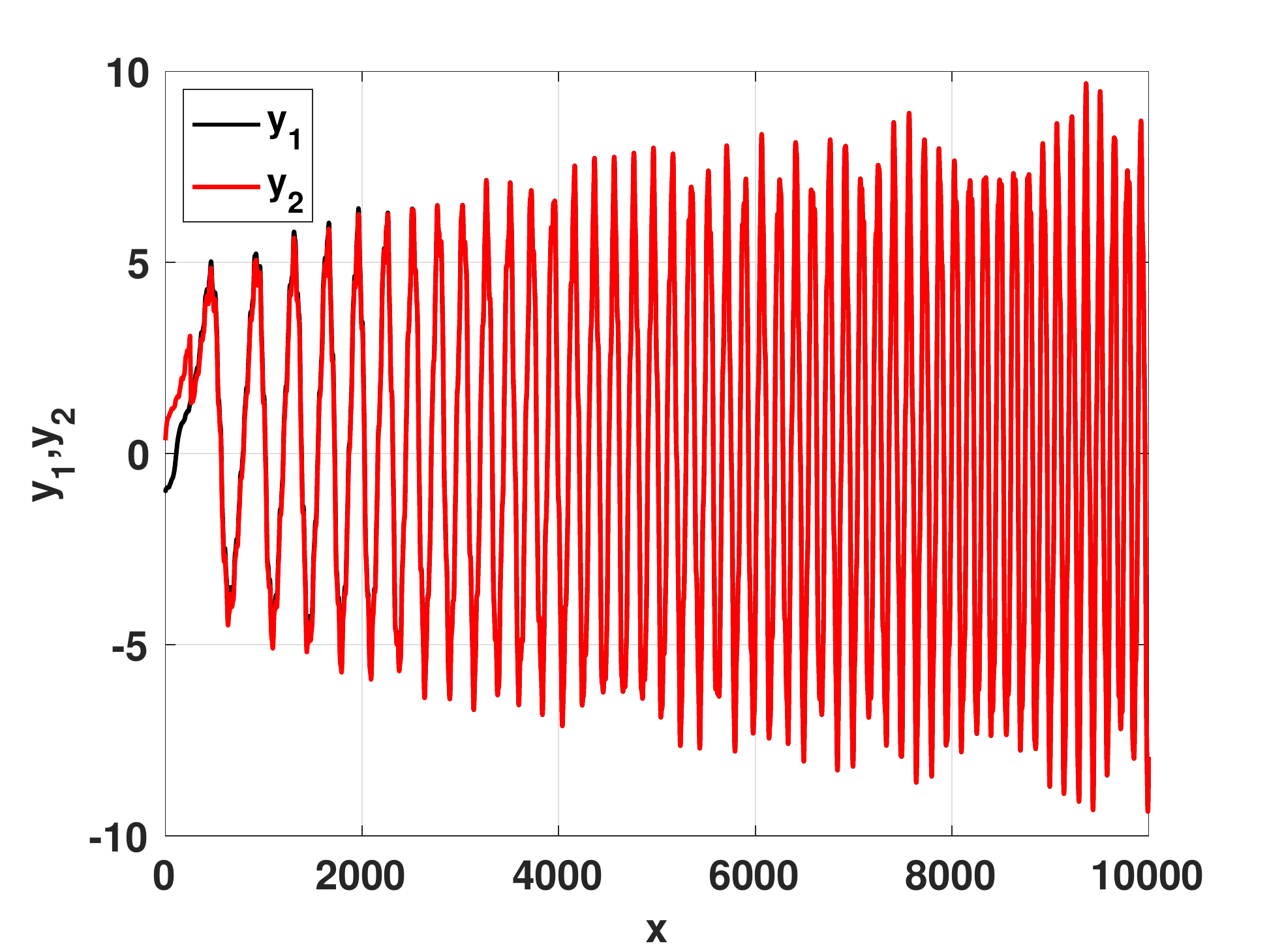}  }
 	\begin{center} (a) \hspace{6cm} (b) \end{center}
 	\caption{(Colour online) Master and slave system synchronization for chaotic BEC sytems: (a) system outputs $x_{1}$ and $x_{2}$, (b)  system outputs $y_{1}$ and $y_{2}$. }
 	\label{fig5}
 \end{figure}

\begin{figure}[h!]
\centerline{\includegraphics[scale=0.40]{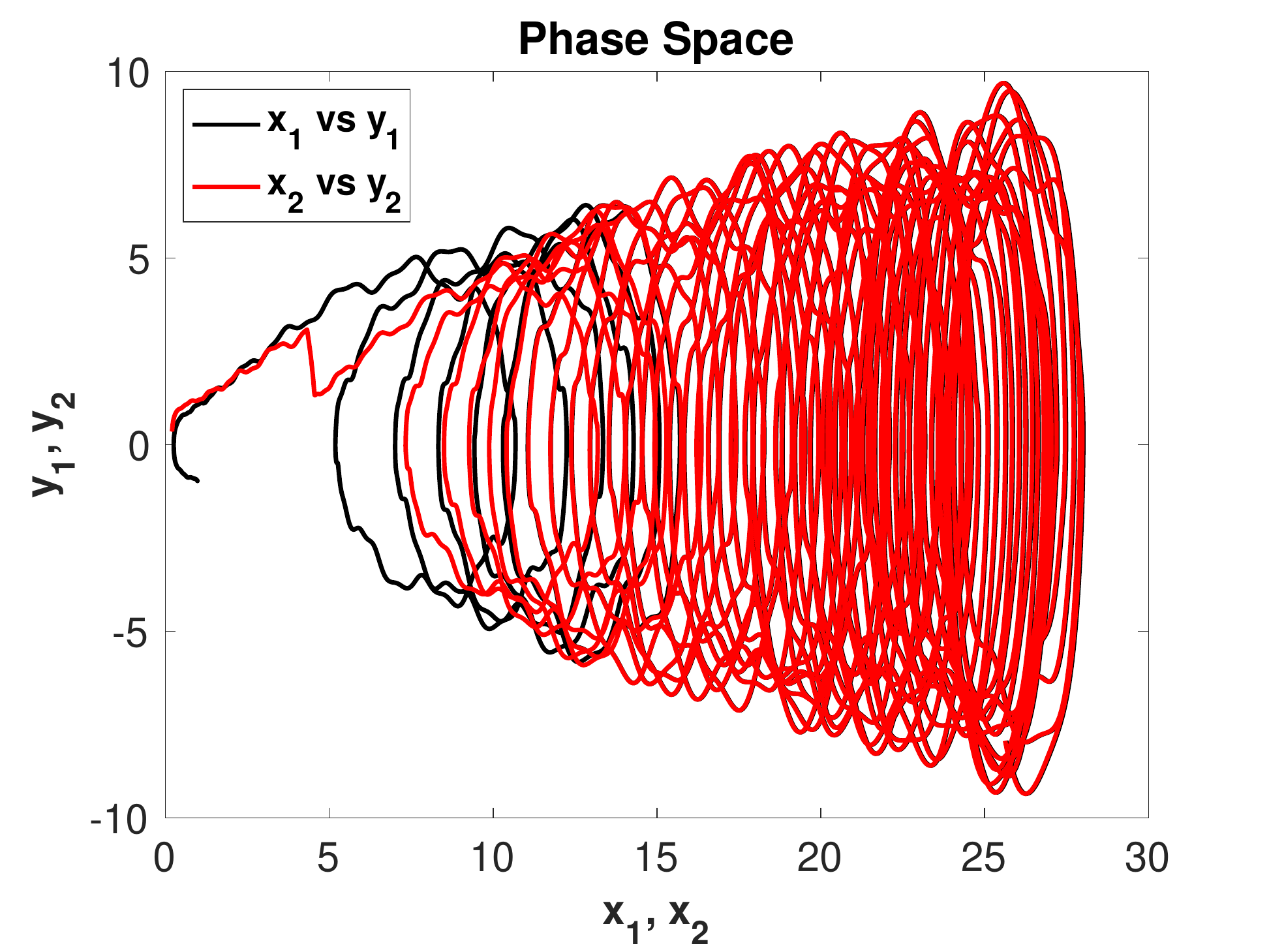}     }
\caption{(Colour online) Synchronized phase space of the master and slave systems. }
\label{fig6}
\end{figure}

\begin{figure}[h!]
\centerline{\includegraphics[scale=0.40]{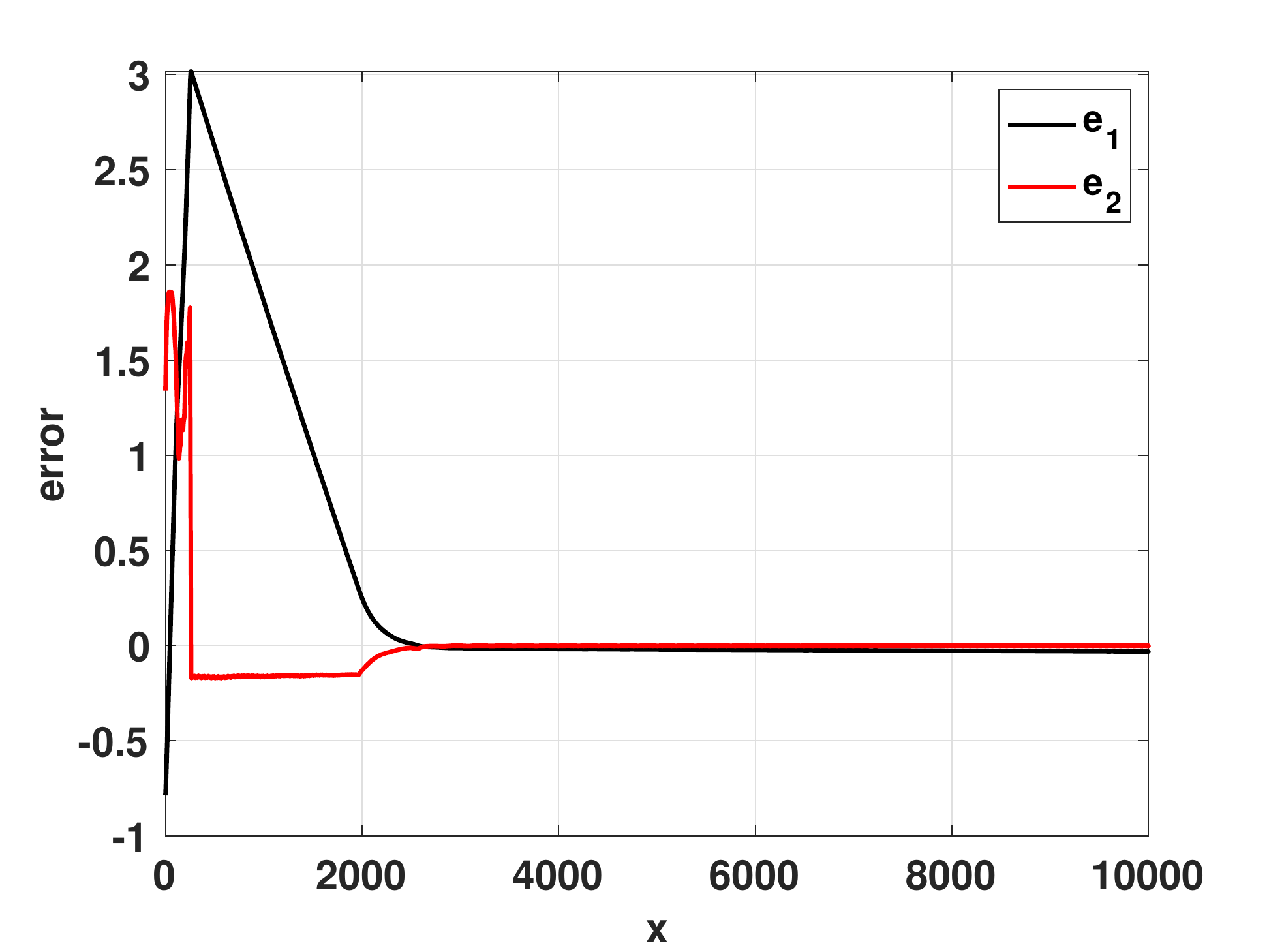}     }
\caption{(Colour online) Error states for master and slave systems. }
\label{fig7}
\end{figure}

\begin{figure}[h!]
\centerline{\includegraphics[scale=0.40]{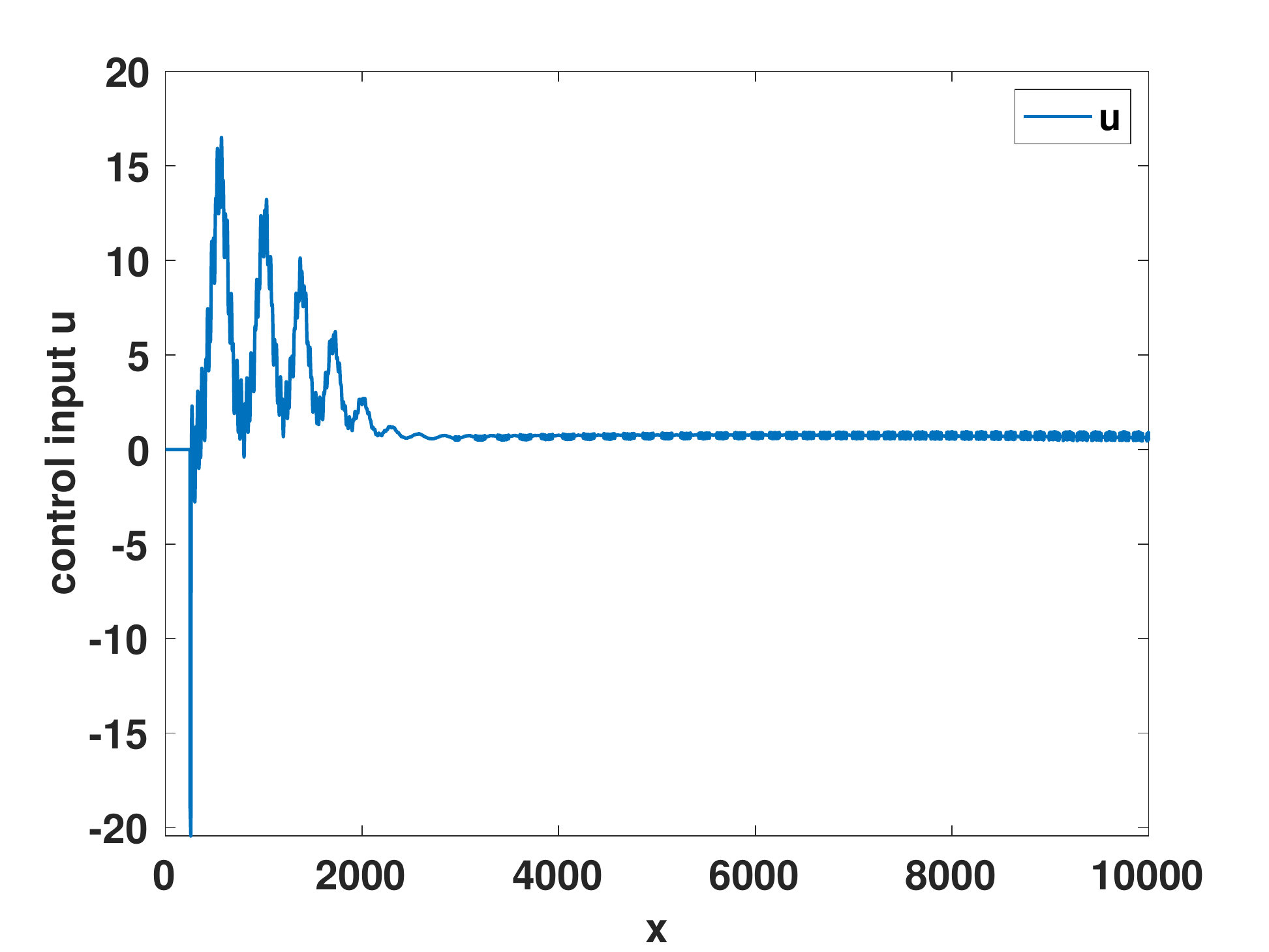}     }
\caption{(Colour online) The evolution of  control input $u$ depending on $x$.}
\label{fig8}
\end{figure}

In this section, the FLC is applied to synchronize the two identical master and slave BEC systems of the form~\eqref{eq11a},~\eqref{eq11b} and~\eqref{eq3_2a},~\eqref{eq3_2b}. For this purpose, the equation systems are solved numerically by using fourth order Runge-Kutta method.  {The} same parameter sets  {are used} in both master and slave system as $J=0.4$, $\nu_{1}=1$, $\nu_{2}=0.8$, $\omega_{1}=2\piup$, $\omega_{2}=5\piup$, $\Gamma=0.1$, $\eta=-0.015$, $\gamma=0.5$. Initial conditions are selected as $(x_1(0),y_1(0))=(1,-1)$ and $(x_2(0),y_2(0))=(0.2,0.3)$ for master and slave systems, respectively. The system  {is evaluated} for $10000$ steps with a step size of $0.01$  {which fulfills the 1D 100 lattice sites boundary condition.} Spatial evaluation and phase space displays for master and slave BEC systems are given in Figure~\ref{fig5} and~\ref{fig6}, respectively.
To observe the effectiveness of the proposed control scheme more clearly, for the first 250 steps, the control input of the fuzzy logic controller, $u_{L}$, is set to zero, such that the master and slave systems are not synchronized.  {Following the $250^{\text{th}}$} step, the computed value of $u_{L}$ is directly fed to the system, which leads the error states $e_{1}$ and $e_{2}$, to converge to zero exponentially. Thus, it can be concluded that two identical BEC systems are synchronized along the flow. The results of error states are given in figure~\ref{fig7}, which  are compatible with spatial evolution and  {the} phase space results given in figure~\ref{fig5} and~\ref{fig6}, respectively. Finally, the graph of the control input $u_{L}$ is given in figure~\ref{fig8}, from which it can be inferred that $u_{L}$ is zero until $x=250$, and it is computed with regard to equation~\eqref{eq3_2bc} only after  {this} step.

\section{Conclusion}
\label{sec6}
In this paper, the effectiveness of the fuzzy logic controller method to synchronize chaotic BEC systems is theoretically and numerically demonstrated. The chaotic synchronization system consists of the master and slave systems. In  {the} master slave scheme,  {the} given BEC  system is considered as a master system, and  {the} other identical  BEC  system  {is} considered as a slave system.  {The difference between the states of the master and slave system and its derivative is selected as the input signals for the fuzzy logic controller.} The consequent part of the fuzzy rules, which consists of analytical functions of the error and its derivative, is determined using  {the} Lyapunov stability theorem to ensure the stability of the synchronization process. Furthermore, this way the dependence of the algorithm  {on} the completeness and accuracy of the expert knowledge is also  {overcome}. In addition, the numerical simulation results show that the error dynamics of the identical chaotic BEC synchronization systems are regulated to zero asymptotically in shorter time in spite of the overall system undergoing irregularity.

 {The chaos which has a destructive role for BEC could disrupt the stability in the condensate.} Therefore, controlling the chaos is of great importance for the creation of BEC.  {A} distinguishing feature of  {the present}  work is the  first synchronization of the identical chaotic Bose-Einstein condensate held in a 1D tilted bichromatical optical lattice potential by using  {the} fuzzy logic control technique.  {Consequently}, this study will make a significant contribution in this  {field.}


\begin{thebibliography}{99}

\bibitem{Work2} Bose S., Z. Phys., 1924, \textbf{26}, 178--181, \doi{10.1007/BF01327326}.

\bibitem{Work3} Einstein A., In: Akademie-Vortr{\"a}ge. Sitzungsberichte der Preu{\ss}ischen Akademie der Wissenschaften 1914-1932, Simon D. (Ed.), 2005, 245--257.

\bibitem{Anderson1} Anderson~M.~H., Ensher~J.~R., Matthews~M.~R., Wieman~C.~E., Cornell~E. A., Science, 1995, \textbf{269}, No.~5221, 198--201, \doi{10.1126/science.269.5221.198}.

\bibitem{Davis1} Davis~K.~B., Mewes~M.~O., Andrews~M.~R., van~Druten~N.~J., Durfee~D.~S., Kurn~D.~M., Ketterle~W., Phys. Rev. Lett., 1995, \textbf{75}, 3969--3973, \doi{10.1103/PhysRevLett.75.3969}.

\bibitem{KETTERLE1996181} Ketterle~W., Van~Druten~N.~J., Adv. At. Mol. Opt. Phys., 1996, \textbf{37}, 181--236,\\ \doi{10.1016/S1049-250X(08)60101-9}.

\bibitem{WorkC} Ensher J.~R., Jin~D.~S., Matthews~M.~R., Wieman~C.~E., Cornell~E.~A., Phys. Rev. Lett., 1996, \textbf{77}, 4984--4987,\\ \doi{10.1103/PhysRevLett.77.4984}.

\bibitem{Work9} Gross E. P., Phys. Rev. Lett., 1961, \textbf{20}, 454--477, \doi{10.1007/BF02731494}.

\bibitem{Work10} Pitaevskii L., Stringari~S., Bose-Einstein Condensation, Clarendon Press, Oxford, 2003.

\bibitem{eren1} Tosyali~E., Aydogmus F., Yilmaz A., 	Int. J. Mod. Phys. B, 2018, \textbf{32}, No.~23, 1850254,\\ \doi{10.1142/S0217979218502545}.

\bibitem{eren2} Tosyali~E., 	Fluctuation Noise Lett., 2018, \textbf{17}, No.~03, 1850027, \doi{10.1142/S021947751850027X}.

\bibitem{eren3} Tosyali~E., Aydogmus F., Condens. Matter Phys., 2020, \textbf{23}, No.~1, 13001, \doi{10.5488/CMP.23.13001}.

\bibitem{eren4} Tosyali~E., Aydogmus F., J. Phys.: Conf. Ser., 2018, \textbf{1141}, No.~1, 012124, \doi{10.1088/1742-6596/1141/1/012124}.

\bibitem{eren5} Aydogmus F., Tosyali~E., Int. J. Control, 2022, \textbf{95}, No.~3, 620--625, \doi{10.1080/00207179.2020.1808244}.

\bibitem{idowu2013synchronization} Idowu B. A., Vincent~U.~E., J. Chaos, 2013, \textbf{2013}, 723581, \doi{10.1155/2013/723581}.

\bibitem{guo2017projective} Guo R., Nonlinear Dyn., 2017, \textbf{90}, 53--64, \doi{10.1007/s11071-017-3645-4}.

\bibitem{mobayen2018synchronization} Mobayen S., Tchier F., Asian J. Control, 2018, \textbf{20}, No.~1, 71--85, \doi{10.1002/asjc.1512}.

\bibitem{ZhangZhiYing110503} Zhang Z.~Y., Feng X.~Q., Yao Z. H., Jia H.~Y., 	Chin. Phys. B, 2015, \textbf{24}, No.~11, 110503, \doi{10.1088/1674-1056/24/11/110503}.

\bibitem{chen2012chaotic} Chen D., Zhang R., Ma X., Liu S., 	Nonlinear Dyn., 2012, \textbf{69}, 35--55, \doi{10.1007/s11071-011-0244-7}.

\bibitem{vaidyanathan2015hybrid} Vaidyanathan S., Azar A.T., In: Advances and Applications in Sliding Mode Control systems. Studies in Computational Intelligence, Vol. 576, Azar A.,  Zhu Q. (Eds.), Springer, Cham., 2015, 549--569,\\ \doi{10.1007/978-3-319-11173-5_20}.

\bibitem{yau2008chaos} Yau H.~T., Shieh C.~S., Nonlinear Anal. Real World Appl., 2008, \textbf{9}, No.~4, 1800--1810,\\ \doi{10.1016/j.nonrwa.2007.05.009}.

\bibitem{vaidyanathan2016takagi} Vaidyanathan S., Azar A.~T.,	Int. J. Intell. Eng. Inf., 2016, \textbf{4}, No.~2, 135--150, \doi{10.1504/IJIEI.2016.076699}.

\bibitem{topalov2011neuro} Topalov~A. V., Oniz~Y., Kayacan~E., Kaynak~O., Neurocomputing, 2011, \textbf{74}, No.~11, 1883--1893,\\ \doi{10.1016/j.neucom.2010.07.035}.

\bibitem{khanesar2015direct} Khanesar~M.~A., Oniz~Y., Kaynak~O., Gao~H., IEEE/ASME Trans. Mechatron., 2016, \textbf{21}, No.~1, 205--213,\\ \doi{10.1109/TMECH.2015.2498169}.

\bibitem{tirkolaee2020novel} Tirkolaee~E.~B., Mardani~A., Dashtian~Z., Soltani~M., Weber G.~W., J. Cleaner Prod., 2020, \textbf{250}, 119517,\\ \doi{10.1016/j.jclepro.2019.119517}.

\bibitem{blanco2017fuzzy} Blanco-Mesa~F., Merig\'{o}~J., Gil-Lafuente~A.~M., J. Intell. Fuzzy Syst., 2017, \textbf{32}, No.~3, 2033--2050,\\ \doi{10.3233/JIFS-161640}.

\bibitem{kalaiarassan2018one} Kalaiarassan G.,  Somanadh K. M., Thirumalai C., Kumar M. S., Mater. Today: Proc., 2018, \textbf{5}, No.~5, 13547--13555, \doi{10.1016/j.matpr.2018.02.350}.

\bibitem{saez2014fuzzy} S\'{a}ez D., \'{A}vila F., Olivares D., Ca\~{n}izares C., Mar\'{i}n L., IEEE Trans. Smart Grid, 2015, \textbf{6}, No.~2, 548--556,\\ \doi{10.1109/TSG.2014.2377178}.

\bibitem{cheng2016fuzzy} Cheng S. H., Chen S. M., Jian W. S., Inf. Sci., 2016, \textbf{327}, 272--287, \doi{10.1016/j.ins.2015.08.024}.

\bibitem{atsalakis2019bitcoin} Atsalakis G. S., Atsalaki I. G., Pasiouras F., Zopounidis C., Eur. J. Oper. Res., 2019, \textbf{276}, No.~2, 770--780,\\ \doi{10.1016/j.ejor.2019.01.040}.

\bibitem{kostikova2016expert} Kostikova A. V., Tereliansky P. V., Shuvaev A. V., Parakhina V. N., Timoshenko P. N., 	ARPN J. Eng. Appl. Sci., 2016, \textbf{11}, No.~17, 10222--10230, URL \url{http://www.arpnjournals.org/jeas/research_papers/rp_2016/jeas_0916_4906.pdf}.

\bibitem{aghbashlo2017fuzzy} Aghbashlo M., Hosseinpour S., Tabatabaei M., Dadak A., Energy, 2017, \textbf{132}, 65--78,\\ \doi{10.1016/j.energy.2017.05.041}.

\bibitem{bulut2019fuzzy} Bulut G. G., G\"{u}ler H., In: 2019 1st Global Power, Energy and Communication Conference (GPECOM), IEEE, 2019, 30--34, \doi{10.1109/GPECOM.2019.8778568}.

\bibitem{chou2013fuzzy} Chou H. G., Chuang C. F., Wang W. J., Lin J. C., IEEE Trans. Inf. Forensics Secur., 2013, \textbf{8}, No.~12, 2177--2185,\\ \doi{10.1109/TIFS.2013.2286268}.

\bibitem{wang2020fuzzy} Wang R., Zhang Y., Chen Y., Chen X., Xi L., Nonlinear Dyn., 2020, \textbf{100}, 1275--1287, \doi{10.1007/s11071-020-05574-x}.

\bibitem{kuo2007design} Kuo C. L., Int. J. Nonlinear Sci. Numer. Simul., 2007, \textbf{8}, No.~4, 631--636, \doi{10.1515/IJNSNS.2007.8.4.631}.

\bibitem{neciu1991} Nenciu G., Int. J. Nonlinear Sci. Numer. Simul., 1991, \textbf{63}, No.~1, 91, \doi{10.1103/RevModPhys.63.91}.

\bibitem{bond_2003} Buchleitner A., Kolovsky A.~R., Phys. Rev. Lett., 2003, \textbf{91}, No.~25, 253002, \doi{10.1103/PhysRevLett.91.253002}.

\bibitem{fallani2004} Fallani L., De Sarlo L., Lye J. E., Modugno M., Saers R., Fort C., Inguscio M., Phys. Rev. Lett., 2004, \textbf{93}, No.~14, 140406, \doi{10.1103/PhysRevLett.93.140406}.

\bibitem{Denschlag_2002} Denschlag J. H., Simsarian J. E., H\'{a}ffner H., McKenzie C., Browaeys A., Cho D., Helmerson K., J. Phys. B: At. Mol. Opt. Phys., 2002, \textbf{35}, No.~14, 3095--3110, \doi{10.1088/0953-4075/35/14/307}.

\bibitem{FANG200561} Fang J., Hai W., Physica B, 2005, \textbf{370}, No.~1--4, 61--72, \doi{10.1016/j.physb.2005.08.033}.

\bibitem{chua2006} Chua V., Porter M. A., Int. J. Bifurcation Chaos, 2006, \textbf{16}, No.~04, 945--959, \doi{10.1142/S0218127406015222}.

\bibitem{hai2009} Hai W., Zhu Q., Rong S., Phys. Rev. A, 2009, \textbf{79}, No.~2, 023603 , \doi{10.1103/PhysRevA.79.023603}.

\bibitem{Work31} Slotine J. J. E., Li W., Applied Nonlinear Control, Prentice Hall International, London, UK, 1991.  
	
\end{thebibliography}

\ukrainianpart

\title{Синхронізація хаосу в системі БЕК з використанням нечіткого логічного контролера%
}
\author{Е. Тосіалі\refaddr{label1},
	Й. Оніз\refaddr{label2}, Ф. Айдогмуз\refaddr{label3}}
\addresses{
	\addr{label1} Відділення оптики, Професійно-технічне училище охорони здоров'я,  Стамбульський університет Білгі, Кустепе, Сіслі, Стамбул, 34387, Туреччина
	\addr{label2} Кафедра мехатроніки, Факультет інжинерії та природничних наук, Стамбульський університет Білгі, Ейюп, Стамбул, 34060, Туреччина
	\addr{label3} Факультет природничих наук та фізики, Стамбульський університет, Везнечилер, Стамбул, 34134, Туреччина
}

\makeukrtitle
\begin{abstract}
Оскільки наявність хаосу в Бозе-Ейнштейнівському конденсаті (БЕК) відіграє деструктивну роль та може зменшувати стабільність конденсату, контролювання хаоcу має величезне значення для створення БЕК. У цій статті запропонований нечіткий логічний контролер для синхронізації хаотичної динаміки двох ідентичних керуючих рівнянь БЕК систем. На відміну від традиційних підходів, де експертні знання використовується для отримання правил нечіткого контролю для відповідних функцій, у цій роботі згадані правила побудовані з використанням теореми стійкості Ляпунова для забезпечення процесу синхронізації. Ефек\-тивність запропонованого процесу контролю продемонстровано чисельно.
	
	\keywords{нечіткий логічний контролер, синхронізація, хаос, Бозе-Ейнштейнівський конденсат}
	
\end{abstract}

\lastpage
\end{document}